\documentclass{article}

\PassOptionsToPackage{numbers, compress}{natbib}
\usepackage[preprint]{neurips_2023}

\usepackage[utf8]{inputenc} % allow utf-8 input
\usepackage[T1]{fontenc}    % use 8-bit T1 fonts
\usepackage{hyperref}       % hyperlinks
\usepackage{url}            % simple URL typesetting
\usepackage{booktabs}       % professional-quality tables
\usepackage{amsfonts}       % blackboard math symbols
\usepackage{nicefrac}       % compact symbols for 1/2, etc.
\usepackage{microtype}      % microtypography
\usepackage{xcolor}         % colors
\usepackage{graphicx}
\usepackage{subfig}
\usepackage{booktabs}
\usepackage{natbib}
\usepackage{float}
\usepackage{booktabs}
\usepackage{tikz}
\usepackage{collcell}
\usepackage{hyperref}
\usepackage{wrapfig,lipsum,booktabs}
\usepackage{multirow}
\usepackage{braket}

\def\eqref#1{equation~\ref{#1}}
\def\1{\bm{1}}

\usepackage{amsmath}
\usepackage{amssymb}
\usepackage{tikz}
\definecolor{gold}{RGB}{221, 196, 65}
\definecolor{silver}{RGB}{215, 215, 215}
\definecolor{bronze}{RGB}{205, 127, 50}
\newcommand{\tikzcircle}[2][red,fill=red]{\tikz[baseline=-0.7ex]\draw[#1,radius=#2] (0,0) circle ;}%
\newcommand{\first}[1]{\colorbox{gold}{\textbf{#1}}}
\newcommand{\second}[1]{\colorbox{silver}{#1}}
\newcommand{\third}[1]{\colorbox{bronze}{{#1}}}
\definecolor{purpleheart}{rgb}{0.41, 0.21, 0.61}

\title{Neural Embeddings for Protein Graphs}

\author{%
    Francesco ~Ceccarelli\\
    University of Cambridge\\
    \texttt{fc485@cam.ac.uk} \\
  \And
    Lorenzo ~Giusti\thanks{The work was performed while the author was in Cambridge.} \\
  Sapienza University\\
  \texttt{loreanzo.giusti@uniroma1.it} \\
  \AND
    Sean B. ~Holden\\
  University of Cambridge\\
  \texttt{sbh11@cl.cam.ac.uk} \\
  \And
    Pietro ~Li\`{o}\\
    University of Cambridge\\
    \texttt{pl219@cam.ac.uk} \\
}

\begin{document}

\maketitle

\begin{abstract}
Proteins perform much of the work in living organisms, and consequently the development of efficient computational methods for protein representation is essential for advancing large-scale biological research. Most current approaches struggle to efficiently integrate the wealth of information contained in the protein sequence and structure. In this paper, we propose a novel framework for embedding protein graphs in geometric vector spaces, by learning an encoder function that preserves the structural distance between protein graphs. Utilizing Graph Neural Networks (GNNs) and Large Language Models (LLMs), the proposed framework generates structure- and sequence-aware protein representations. We demonstrate that our embeddings are successful in the task of comparing protein structures, while providing a significant speed-up compared to traditional approaches based on structural alignment. Our framework achieves remarkable results in the task of protein structure classification; in particular, when compared to other work, the proposed method shows an average F1-Score improvement of 26\% on out-of-distribution (OOD) samples and of 32\% when tested on samples coming from the same distribution as the training data. Our approach finds applications in areas such as drug prioritization, drug re-purposing, disease sub-type analysis and elsewhere.
\end{abstract}

\section{Introduction}

Proteins are organic macro-molecules made up of twenty standard amino acids. Almost all interactions and reactions which occur in living organisms, from signal transduction, gene transcription and immune function to catalysis of chemical reactions, involve proteins~\cite{morris2022uncovering}. The comparison of proteins and their structures is an essential task in bioinformatics, providing support for protein structure prediction~\cite{kryshtafovych2019critical}, the study of protein-protein docking~\cite{lensink2018challenge}, structure-based protein function prediction~\cite{gherardini2008structure} and many further tasks. Considering the large quantity of protein data stored in the Protein Data Bank (PDB)~\cite{berman2003announcing} and the rapid development of methods for performing protein structure prediction (for example, AlphaFold2~\cite{senior2020improved}), it is desirable to develop methods capable of efficiently comparing the tertiary structures of proteins. 

Generally, protein comparison methods can be divided into two classes: alignment-based methods~\cite{akdel2020caretta, shindyalov1998protein, kihara2003pdb} and alignment-free methods~\cite{xia2022fast, rogen2003automatic, budowski2010fragbag, zotenko2006secondary}. The former aim at finding the optimal structural superposition of two proteins. A scoring function is then used to measure the distance between each pair of superimposed residues. For such methods (for example~\cite{holm1993protein, zhang2005tm}) the superposition of the atomic structures is the main bottleneck as it has been proven to be an NP-hard problem~\cite{lathrop1994protein}. On the other hand, alignment-free methods try to represent each protein in the form of a descriptor, and then to measure the distance between pairs of descriptors~\cite{xia2022fast}. Descriptors need to satisfy two requirements: 1) their size should be fixed and independent of the length of proteins; 2) they should be invariant to rotation and translation of proteins. 

The template modeling score (TM-score)~\cite{zhang2004scoring} is a widely used metric for assessing the structural similarity between two proteins. It is based on the root-mean-square deviation (RMSD) of the atomic positions in the proteins, but considers the lengths of the proteins and the number of residues that can be superimposed. TM-score has been shown to be highly correlated with the similarity of protein structures and can be used to identify structurally similar proteins, even when they have low sequence similarity. Unfortunately, computing TM-scores is computationally intractable even for relatively small numbers of proteins. TM-align~\cite{zhang2005tm}, one of the popular alignment-based methods, takes about 0.5 seconds for one structural alignment on a 1.26 GHz PIII processor. As such, computing TM-scores for existing databases, containing data for millions of proteins, is unaffordable. While several deep learning methods for protein comparison have been developed (for example, DeepFold~\cite{liu2018learning} and GraSR~\cite{xia2022fast}) they suffer from major drawbacks: 1) they are trained by framing the protein comparison task as a classification problem---that is, predicting if two proteins are structurally similar---and hence fail to directly incorporate TM-scores in the loss function formulation; 2) they produce latent representations (embeddings) which do not integrate the information contained in the protein sequences and structures; 3) they usually do not exploit the inductive bias induced by the topology of graph-structured proteins, and they fail to consider different geometries of the latent space to match well the underlying data distribution. 

In this paper, we address the aforementioned limitations of current protein embedding methods by proposing an efficient and accurate technique that integrates both protein sequence and structure information. In detail, we first construct protein graphs where each node represents an amino acid in the protein sequence. We then generate features for each amino acid (node in the graph) using Large Language Models (LLMs) before applying Graph Neural Networks (GNNs) to embed the protein graphs in geometric vector spaces while combining structural and sequence information. By incorporating TM-scores in the formulation of the loss function, the trained graph models are able to learn a mapping that preserves the distance between the input protein graphs, providing a way to quickly compute similarities for every pair of unseen proteins. We evaluated the proposed approach and its ability to generate meaningful embeddings for downstream tasks on two protein datasets. On both, the proposed approach reached good results, outperforming other current state-of-the-art methods on the task of structural classification of proteins on the SCOPe dataset~\cite{fox2014scope}.

\textbf{Contribution} The main contributions of this paper can be summarised as follow: \textit{(i)} A novel learning framework for generating protein representations in geometric vector spaces by merging structural and sequence information using GNNs and LLMs. \textit{(ii)} A quick and efficient method for similarity computation between any pair of proteins. \textit{(iii)} An evaluation of the ability of our embeddings, in both supervised and unsupervised settings, to solve downstream protein classification tasks, and a demonstration of their superior performance when compared to current state-of-the-art methods. Our approach finds a plethora of applications in the fields of bioinformatics and drug discovery.

\section{Background and Related Work}

Several alignment-based methods have been proposed over the years, each exploiting different heuristics to speed up the alignment process. For example, in DALI~\cite{holm1999using}, Monte Carlo optimization is used to search for the best structural alignment. In~\cite{shindyalov1998protein} the authors proposed combinatorial extension (CE) for similarity evaluation and path extension. An iterative heuristic based on the Needleman–Wunsch dynamic programming algorithm~\cite{needleman1970general} is employed in TM-align~\cite{zhang2005tm}, SAL~\cite{krishna1997pdb} and STRUCTAL~\cite{zhang1997unified}. Examples of alignment-free approaches are Scaled Gauss Metric (SGM)~\cite{rogen2003automatic} and the Secondary Structure Element Footprint (SSEF)~\cite{zotenko2006secondary}. SGM treats the protein backbone as a space curve to construct a geometric measure of the conformation of a protein, and then uses this measure to provide a distance between protein shapes. SSEF splits the protein into short consecutive fragments and then uses these fragments to produce a vector representation of the protein structure as a whole. More recently, methods based on deep learning have been developed for the task of protein structure comparison. For instance, DeepFold~\cite{liu2018learning} used a deep convolutional neural network model trained with the max-margin ranking loss function~\cite{wang2016learning} to extract structural motif features of a protein, and learn a fingerprint representation for each protein.  Cosine similarity was then used to measure the similiarity scores between proteins. DeepFold has a large number of parameters, and fails to exploit the sequence information and the topology of graph-structured data. GraSR~\cite{xia2022fast} employs a contrastive learning framework, GNNs and a raw node feature extraction method to perform protein comparison. Compared to GraSR, we present a general framework to produce representations of protein graphs where the distance in the embedding space is correlated with the structural distance measured by TM-scores between graphs. Finally, our approach extends the work presented in~\cite{corso2021neural}, which was limited to biological sequence embeddings, to the realm of graph-structured data.

\section{Material and Methods}

\begin{figure}[h]
    \centering
    \includegraphics[width=1\textwidth]{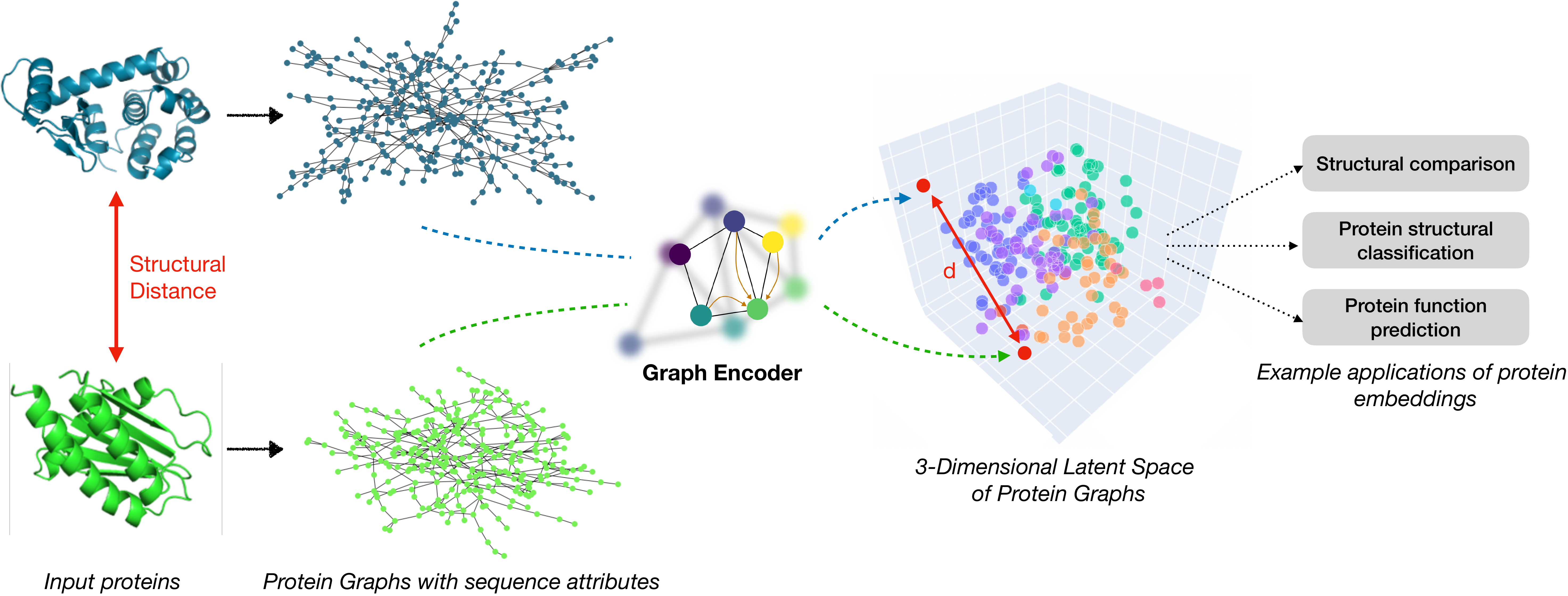}
    \caption{We learn an encoder function that preserves the structural distance, measured by the TM-score, between two input proteins. We construct protein graphs by combining sequence and structure information as shown in Figure \ref{fig:graph_construction}. A distance function $d$ defines the shape of the latent space. The generated embeddings can be used for a variety of applications in bioinformatics and drug discovery. (For simplicity, this Figure depicts a 3-dimensional latent space.)}
    \label{fig:pipeline_fig}
\end{figure}

The core approach, shown in Figure~\ref{fig:pipeline_fig}, 
is to map graphs into a continuous space so that the distance between embedded points reflects the distance between the original graphs measured by the TM-scores. The main components of the proposed framework are the geometry of the latent space, a graph encoder model, a sequence encoder model, and a loss function. Details for each are as follows.

\paragraph{Latent space geometry} 
The distance function used ($d$ in Figure~\ref{fig:pipeline_fig}) defines the geometry of the latent space into which embeddings are projected. In this work we provide a comparison between Euclidean, Manhattan, Cosine and squared Euclidean (referred to as Square) distances (details in Appendix~\ref{DF}).

\paragraph{Graph encoder model}
The encoder performs the task of mapping the input graphs to the embedding space. A variety of models exist for this task, including linear, Multi-layer Perceptron (MLP), LSTM~\cite{cho2014learning}, CNN~\cite{fukushima1980neocognitron} and Transformers~\cite{vaswani2017attention}. Given the natural representation of proteins as graphs, we chose GNNs as encoder models. We have constructed the molecular graphs of proteins starting from PDB files. A PDB file contains structural information such as 3D atomic coordinates. Let $G=(V,E)$ be a graph representing a protein, where each node $v \in V$ is a residue and interaction between the residues is described by an edge $e \in E$. Two residues are connected if they have any pair of atoms (one from each residue) separated by a Euclidean distance less than a threshold distance. The typical cut-off, which we adopt in this work, is 6 angstroms (Å)~\cite{chen2021structure}.

\begin{figure}[]
    \centering
    \includegraphics[width=0.85\textwidth]{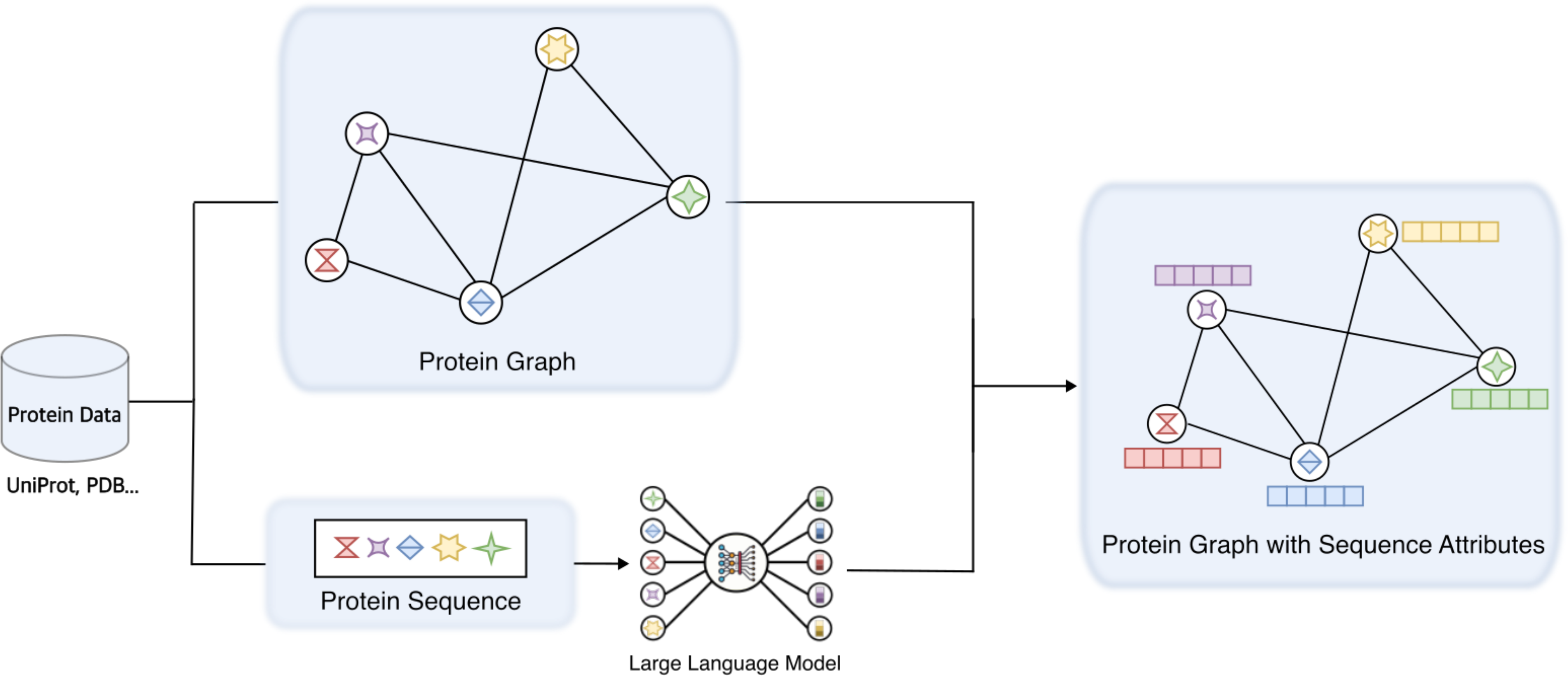}
    \caption{Graph representation of a protein, which combines sequence and structure. Starting from protein data (a PDB file from, for example, UniProt~\cite{uniprot2023uniprot} or PDB~\cite{berman2003announcing}), we extract protein sequence and structure information. We construct graphs where each node represents an amino acid in the protein sequence. We then generate features for each node in the graph using Large Language Models pre-trained on protein sequences.}
    \label{fig:graph_construction}
\end{figure}

\paragraph{Sequence encoder model}
Given the graph representation of a protein, each node $v$ of the graph (each residue) must be associated with a feature vector. Typically, features extracted from protein sequences by means of LLMs have exhibited superior performances compared to handcrafted features. We experimented with five different sequence encoding methods: 1) a simple one-hot encoding of each residue in the graph, 2) seven physicochemical properties of residues as extracted by~\cite{meiler2001generation}, which are assumed to influence the interactions between proteins by creating hydrophobic forces or hydrogen bonds between them, 3) the BLOcks SUbstitution Matrix (BLOSUM)~\cite{henikoff1992amino}, which counts the relative frequencies of amino acids and their substitution probabilities, 4) features extracted from protein sequences employing a pre-trained BERT-based transformer model (ProBert~\cite{brandes2022proteinbert}), and 5) node features extracted using a a pre-trained LSTM-based language model (SeqVec~\cite{heinzinger2019modeling}). Table~\ref{tab:dim_table} 
\begin{table}[]
\centering
\caption{Investigated node attributes and their dimensions. BERT and LSTM features are extracted using LLMs pre-trained on protein sequences (ProBert~\cite{brandes2022proteinbert} and SeqVec~\cite{heinzinger2019modeling}).}
\vspace{3px}
\begin{tabular}{@{}lc@{}}
\toprule
\textbf{Feature}                & \multicolumn{1}{l}{\textbf{Dimension}} \\ \midrule
One hot encoding of amino acids & 20              \\
Physicochemical properties       & 7                                      \\
BLOcks SUbstitution Matrix      & 25                                     \\
BERT-based language mod         & 1024                                   \\
LSTM-based language model      & 1024                                   \\ \bottomrule
\end{tabular}
\label{tab:dim_table}
\end{table}
summarizes the node features and their dimensions, while Figure~\ref{fig:graph_construction} 
depicts the process of constructing a protein graph with node features, starting from the corresponding protein data.

\paragraph{Loss function}
The loss function used, which minimises the MSE between the graph distance and its approximation as the distance between the embeddings, is
\begin{equation}
    L\left(\theta,G\right)=\sum_{g_1,g_2\in G}\left(\text{TM}\left(g_1,g_2\right)-d\left(\text{GNN}_\theta\left(g_1\right), \text{GNN}_\theta\left(g_2\right)\right)\right)^2
\label{eq:loss}
\end{equation}
where $G$ is the training set of protein graphs and $\text{GNN}_\theta$ is the graph encoder. The TM-score is a similarity metric in the range (0,1], where 1 indicates a perfect match between two structures. Since the formulation of the loss is expressed in terms of distances, we reformulate the TM-scores as a distance metric by simply computing  $\text{TM}(g_1,g_2) = 1 - \text{TM}_{\text{score}}(g_1,g_2)$. By training neural networks to minimize the loss in Equation~\ref{eq:loss}, we encourage the networks to produce latent representations such that the distance between these representations is proportional to the structural distance between the input graphs.

\section{Protein Datasets}

We evaluated the proposed approach on two protein datasets. First, we downloaded the human proteome from UniProt\footnote{\url{https://www.uniprot.org}} and sub-selected 512 protein kinases. To obtain the TM-scores to train the graph models, we evaluated the structural similarity using TM-align~\cite{zhang2005tm}. All-against-all alignment yielded a dataset composed of 130,816 total comparisons. Every kinase in the dataset is categorized in one of seven family groups: a) AGC (63 proteins), b) CAMK (82 proteins), c) CK1 (12 proteins), d) CMGC (63 proteins), e) STE (48 proteins), f) TK (94 proteins), and g) TKL (43 proteins). The number of nodes in the graphs ranges from 253 to 2644, with an average size of approximately 780 nodes. The average degree in the graphs is approximately 204, the average diameter of the graphs is approximately 53 nodes and the maximum diameter is 227 nodes. We further used SCOPe v2.07 (March 2018) as a benchmark dataset~\cite{fox2014scope}. This dataset contains 13,265 protein domains classified in one of seven classes: a) all alpha proteins (2286 domains), b) all beta proteins (2757 domains), c) alpha and beta proteins (a/b) (4148 domains), d) alpha and beta proteins (a+b) (3378 domains), e) multi-domain proteins (alpha and beta) (279 domains), f) membrane and cell surface proteins and peptides (213 domains), and g) small proteins (204 domains). We again used TM-align with all-against-all settings to construct a dataset of approximately 170 millions comparisons. To reduce the computational time and cost during training, we randomly sub-sampled 100 comparisons for each protein to create a final dataset of 1,326,500 comparisons. For this dataset, the number of nodes in the graphs ranges from 30 to 9800, with an average size of approximately 1978 nodes. The average degree is approximately 90, the average diameter of the graphs is approximately 9 nodes and the maximum diameter is 53 nodes. Compared to benchmark graph datasets (for example~\cite{ZINCdataset} and~\cite{dwivedi2022long}) we evaluated our approach on graphs of significantly larger size 
(84 and 13 times more nodes than the molecular graphs in ~\cite{ZINCdataset} and ~\cite{dwivedi2022long}, respectively).

\section{Experimental Results}

\paragraph{Experimental settings}

We evaluate the proposed framework using Graph Convolutional Networks (GCNs)~\cite{kipf2016semi}, Graph Attention Networks (GATs)~\cite{velivckovic2017graph}, and GraphSAGE~\cite{hamilton2017inductive} (Appendix~\ref{GA}). All the models were implemented with two graph layers in PyTorch geometric~\cite{pyg2019} to learn protein embeddings of size 256. Adam optimizer~\cite{kingma2014adam} with a learning rate of 0.001 was used to train the models for 100 epochs with a patience of 10 epochs. The batch size was set to 100. We used 4 attention heads in the GAT architecture. For each model, Rectified Linear Units (ReLUs)~\cite{nair2010rectified} and Dropout~\cite{srivastava2014dropout} were applied after each layer, and mean pooling was employed as readout function to obtain graph-level embeddings from the learned node-level representations. Finally, each experiment was run with 3 different seeds to provide uncertainty estimates.

\paragraph{Kinase embeddings} 
For the generation of the embeddings, we used 80\% of the kinase proteins for training and the remaining 20\% for testing. Table~\ref{tab:res1_table} 
\begin{table}[]
\centering
\small
\caption{MSE results for different feature types, distance functions and graph encoder models on the kinase dataset. We use gold \tikzcircle[gold,fill=gold]{2pt}, silver \tikzcircle[silver,fill=silver]{2pt}, and bronze \tikzcircle[bronze,fill=bronze]{2pt} colors to indicate the first, second and third best performances, respectively. For each model, the best scores are consistently reached with LSTM-extracted features and Euclidean geometry of the embedding space. Across all models, GAT embeddings exhibit the best performance. For all the models, MSE scores are lower for features extracted by means of LLMs (BERT and LSTM) compared to handcrafted feature extraction methods (one-hot, biochemical and BLOSUM).}
\vspace{3px}
\begin{tabular}{@{}clcccc@{}}
\toprule
\textbf{Model}                                                        & \multicolumn{1}{c}{\textbf{Feature}} & \multicolumn{4}{c}{\textbf{Distance}}   \\
&                  & \texttt{Cosine} & \texttt{Euclidean} & \texttt{Manhattan} & \texttt{Square} \\ \midrule
\multirow{5}{*}{GCN}                                                  & One hot          & $0.0194 \pm 0.002$  & $0.0380  \pm 0.003$  & $0.0192 \pm 0.001$    & $0.0729 \pm 0.004$ \\
& Physicochemical  & $0.0343 \pm 0.012$ & $0.0483 \pm 0.009$    & $0.0397 \pm 0.003$    & $0.1109 \pm 0.007$ \\
& BLOSUM           & $0.0327 \pm 0.071$ & $0.0271 \pm 0.043$    & $0.0450 \pm 0.013$    & $0.0697 \pm 0.023$ \\
& BERT             & $0.0110 \pm 0.003$ & $0.0103 \pm 0.001$    & $0.0131 \pm 0.006$    & $0.0138 \pm 0.009$ \\
& LSTM             & $0.0105 \pm 0.002$ & \third{$0.0088 \pm 0.004$}    & $0.0156 \pm 0.001$    & $0.0107 \pm 0.004$ \\ \midrule
\multirow{5}{*}{GAT}                                                  & One hot          & $0.0171 \pm 0.001$ & $0.0320 \pm 0.012$    & $0.0171 \pm 0.011$    & $0.0758 \pm 0.009$ \\
& Physicochemical  & $0.0295 \pm 0.007$ & $0.0328 \pm 0.006$    & $0.0220 \pm 0.004$    & $0.0856 \pm 0.023$ \\
& BLOSUM           & $0.0245 \pm 0.012$ & $0.0163 \pm 0.009$    & $0.0124 \pm 0.011$    & $0.0307 \pm 0.009$ \\
& BERT             & $0.0091 \pm 0.018$ & $0.0095 \pm 0.008$    & $0.0078 \pm 0.009$    & $0.0133\pm 0.011$ \\
& LSTM             & $0.0088 \pm 0.009$ & \first{$0.0073 \pm 0.004$}    & $0.0086 \pm 0.006$    & $0.0101 \pm 0.009$ \\ \midrule
\multirow{5}{*}{\begin{tabular}[c]{@{}l@{}}Graph\\ SAGE\end{tabular}} & One hot          & $0.0243 \pm 0.002$ & $0.0227 \pm 0.011$    & $0.0156 \pm 0.009$    & $0.0424\pm 0.010$ \\
& Physicochemical  & $0.0301 \pm 0.004$ & $0.0266 \pm 0.008$    & $0.0310 \pm 0.011$    & $0.0578 \pm 0.009$ \\
& BLOSUM           & $0.0285 \pm 0.007$ & $0.0172 \pm 0.008$    & $0.0342 \pm 0.002$    & $0.0368 \pm 0.007$ \\
& BERT             & $0.0097 \pm 0.011$ &$ 0.0089 \pm 0.007$    & $0.0101 \pm 0.007$    & $0.0107 \pm 0.009$ \\
& LSTM             & $0.0093 \pm 0.003$& \second{$0.0084 \pm 0.005$}    & $0.0143 \pm 0.007$    & $0.0094 \pm 0.008$ \\ \bottomrule
\end{tabular}
\vspace{-10pt}
\label{tab:res1_table}
\end{table}
shows the MSE values for the graph encoders, using different choices for distance functions and node features. For each model, the best scores are consistently reached with LSTM-extracted features and Euclidean geometry of the embedding space. Across all models, GAT embeddings exhibit the lowest MSE, followed by GarphSAGE and GCN. From Table~\ref{tab:res1_table}, it is clear that using pre-trained language models to extract node features from protein sequences leads to better results. MSE scores for all distances across all encoder models are lower when using BERT and LSTM features. Furthermore, the LSTM-extracted features perform consistently better compared to the BERT ones. BLOSUM and Physicochemical features are also usually associated with higher MSE for all distances and models, indicating that they are poorly correlated to TM-scores. 

\begin{figure}[h]
    \centering
    \includegraphics[width=1\textwidth]{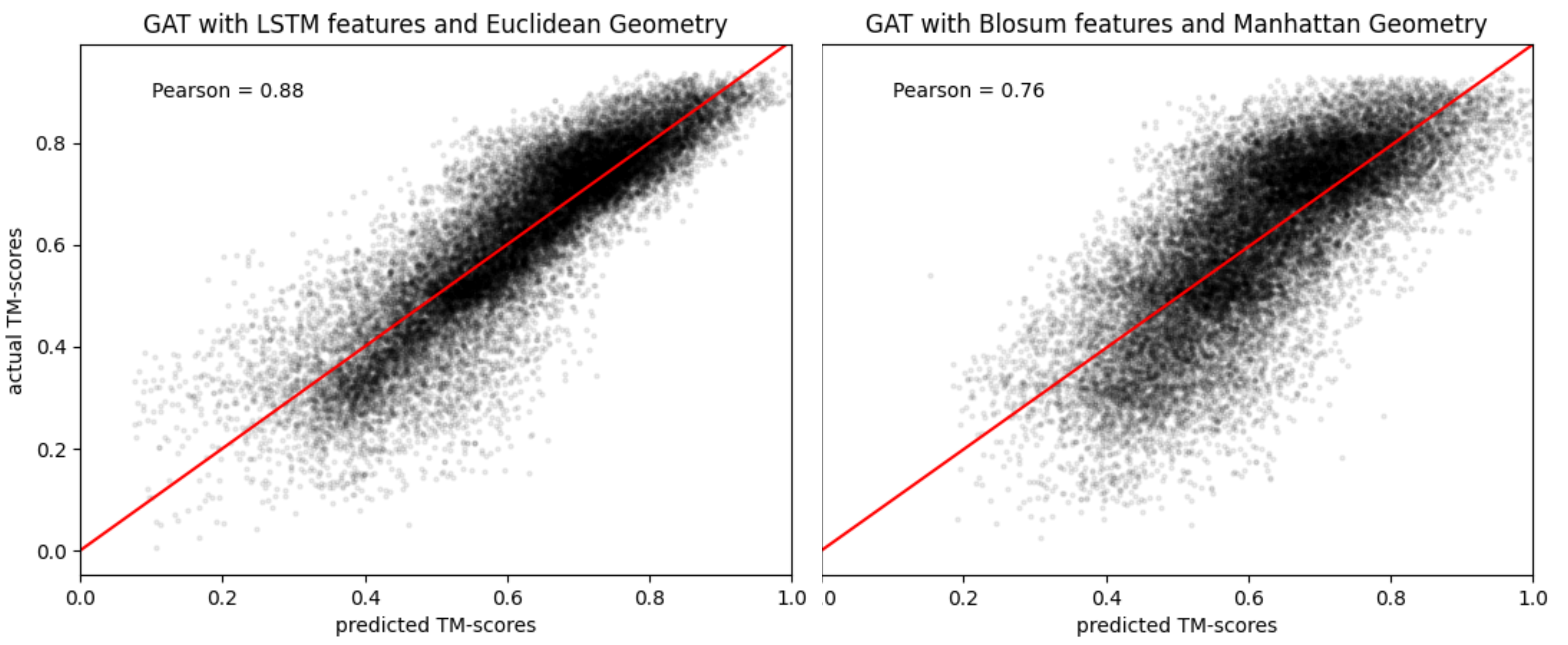}
    \caption{Actual versus predicted TM-scores. Using LSTM features and Euclidean geometry (\textit{left}) results in predictions which follow more tightly the red line of the oracle compared to BLOSUM features in the Manhattan space (\textit{right}).}
    \label{fig:TM_correlation}
\end{figure}

\paragraph{Fast inference of TM-scores} 
We employed the trained GAT architectures from Table~\ref{tab:res1_table} to predict the TM-scores for the kinase pairs in the test set. In Figure~\ref{fig:TM_correlation}, 
we show the predicted versus actual TM-scores for two combinations of features and embedding geometries. The left plot in Figure~\ref{fig:TM_correlation} uses LSTM-extracted features and Euclidean space, while the right one shows predictions for BLOSUM features and Manhattan space. The complete quantitative evaluations, measured by Pearson correlation between model predictions and true TM-scores for all distances and features, are reported in Appendix~\ref{TMAPP}. As in Table~\ref{tab:res1_table}, the best performances are reached when employing LSTM and BERT features while BLOSUM and Physicochemical features lead to the poorest performances (Appendix~\ref{TMAPP}). The highest correlation score, reflecting the results reported in Table~\ref{tab:res1_table}, is reached when employing LSTM features and Euclidean distance (Figure~\ref{fig:TM_correlation}). It is worth noticing that, for the 26,164 comparisons in the test set, the proposed approach took roughly 120 seconds to compute TM-scores. Executing TM-align with the same number of comparisons took 57,659 seconds ($\approx 16$ hours). Details of the TM-score inference times for all the models are given in Appendix~\ref{TimeInference}. {\em The major speed-up provided by performing inference using machine learning models makes the proposed approach applicable to datasets comprising millions of proteins.}

\paragraph{Ablation study: structure removal} 
Coupling GNNs with LLMs provides a means of integrating the information coming from the structure and sequence of proteins. To analyse the benefits of exploiting the topology induced by the graph structures, we performed an ablation study which disregards such information. DeepSet~\cite{zaheer2017deep} considers objective functions defined on sets, that are invariant to permutations. Using a DeepSet formulation, we constructed protein graphs with features where each node is only connected to itself. As for the graph models, we trained DeepSet to minimize the loss function in Equation~\ref{eq:loss} and report the results in Table~\ref{tab:ablation}. 
\begin{table}[]
\centering
\small
\caption{MSE values for an ablation study which disregards the topological information induced by the structure of the protein graphs. We use gold \tikzcircle[gold,fill=gold]{2pt}, silver \tikzcircle[silver,fill=silver]{2pt}, and bronze \tikzcircle[bronze,fill=bronze]{2pt} colors to indicate the first, second and third best performances, respectively. By ignoring the neighborhood and the structural information, the MSEs are significantly higher (p-value of t-test $< 0.05$) compared to GNNs.}
\vspace{2px}
\begin{tabular}{@{}llcccc@{}}
\toprule
\textbf{Model}           & \textbf{Feature} & \multicolumn{4}{c}{\textbf{Distance}}   \\
                         &                  & \texttt{Cosine} & \texttt{Euclidean} & \texttt{Manhattan} & \texttt{Square} \\ \midrule
\multirow{5}{*}{DeepSet} & One Hot          & $0.1742 \pm 0.003$& $0.0421 \pm 0.002$    & $0.0358 \pm 0.001$    & $0.0714 \pm 0.003$ \\
                         & Physicochemical  & $0.1766 \pm 0.010$& $0.0437 \pm 0.006$    & $0.0464 \pm 0.004$   & $0.0900\pm 0.006$ \\
                         & BLOSUM           & $0.1553\pm 0.003$ & $0.0381 \pm 0.009$    & $0.0558 \pm 0.008$    & $0.0914 \pm 0.008$ \\
                         & BERT features    & \third{$0.0132 \pm 0.004$}& \second{$0.0129\pm 0.005$}    & $0.0192\pm 0.005$    & $0.0220\pm 0.004$ \\
                         & LSTM features    & $0.0141 \pm 0.003$&\first{$ 0.0116 \pm 0.010$} & $0.0348\pm 0.006$    & $0.0200\pm 0.007$ \\ \bottomrule
\end{tabular}
\label{tab:ablation}
\end{table}
Similarly to Table~\ref{tab:res1_table}, the best MSE scores are reached when using LSTM features and Euclidean geometry. The scores in Table~\ref{tab:ablation}, computed by disregarding the graph connectivity and neighborhood information, are significantly higher than those reported in Table~\ref{tab:res1_table} (p-value of t-test $< 0.05$ compared to GCN, GAT and GraphSAGE). By considering patterns of local connectivity and structural topology, GNNs are able to learn better protein graph representations compared to models which only exploit sequence-derived features. 

\begin{figure}[]
    \centering
    \includegraphics[width=0.6\textwidth]{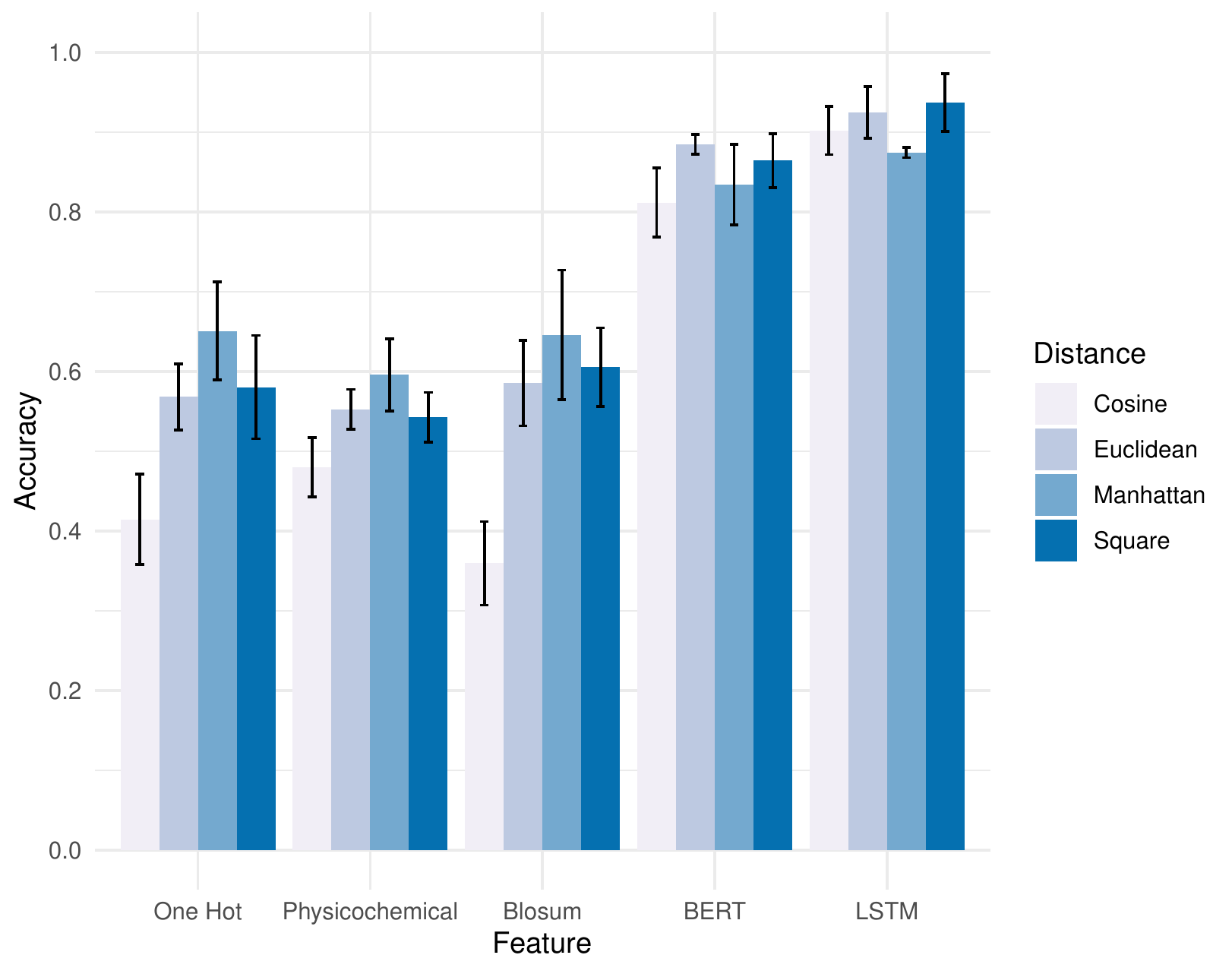}
    \caption{Accuracy of classification for kinase family prediction using the embeddings generated by the GAT models. The highest accuracy value of 93.7\% is reached with LSTM features and the Square distance function.}
    \label{fig:Multiclass_kinases}
\end{figure}

\paragraph{Downstream task of kinase classification} 
To prove the usefulness of the learned embeddings for downstream tasks, we set out to classify each kinase into one of the seven family groups (AGC, CAMK, CK1, CMGC, STE, TK, TKL). Using the embeddings generated by the GAT models, we trained an MLP, composed of 3 layers of size 128, 64 and 32 respectively, and a SoftMax classification head. The accuracy of classification, computed as the average result of 5-fold cross-validation, for each feature type and distance function is reported in Figure~\ref{fig:Multiclass_kinases}. 
The results are consistent with Table~\ref{tab:res1_table}: the best accuracies are obtained when using LSTM- and BERT-extracted sequence features, while handcrafted feature extraction methods (one hot, BLOSUM and physicochemical) provide the poorest performance. The highest accuracy values of 93.7\% and 92.48\% are reached with LSTM features and Square and Euclidean distance functions, respectively.

\paragraph{Embedding out of distribution samples} 
Being able to use pre-trained models for related or similar tasks is essential in machine learning. We tested the ability of the proposed graph models to generalize to new tasks by generating embeddings for the 13,265 proteins in the SCOPe dataset after being trained only on kinase proteins. Given the better performance provided by the use of LSTM features, in this section we constructed protein graphs with LSTM attributes and used a 3-Layer MLP as before to assign the GAT-generated protein embeddings from the SCOPe dataset to the correct class. Results of this evaluation, measured as average F1-score across 5 folds for each distance function, are shown in Table~\ref{tab:tl_1} (F1-Score out of distribution (OOD)). 
\begin{table}[]
\centering
\small
\caption{Out of distribution (OOD) classification results on SCOPe proteins (\textbf{F1-Score (OOD)}). We use gold \tikzcircle[gold,fill=gold]{2pt}, silver \tikzcircle[silver,fill=silver]{2pt}, and bronze \tikzcircle[bronze,fill=bronze]{2pt} colors to indicate the first, second and third best performances, respectively. Despite the different training data, the GAT model with Euclidean and Square geometry outperforms all other approaches trained on SCOPe proteins. Classification results for embeddings generated after training on SCOPe proteins are also shown (\textbf{F1-Score}); in this case, the proposed approach outperforms the others by a larger margin for all choices of latent geometries.}
\vspace{3px}
\begin{tabular}{@{}llcc@{}}
\toprule
\textbf{Model}                 & \textbf{Distance}     & \textbf{F1-Score (OOD)} & \textbf{F1-Score}  \\ \midrule
\multirow{4}{*}{GAT} & \texttt{Cosine}                & $0.6906 \pm 0.0044$     & $0.8290 \pm 0.008$ \\
                               & \texttt{Euclidean}             & \first{$0.8204\pm 0.006$}       & \first{$0.8557 \pm 0.002$} \\
                               & \texttt{Manhattan}             & $0.7055   \pm 0.006$    & \second{$0.8481\pm0.007$}   \\
                               & \texttt{Square}                & \second{$0.8185\pm 0.004$}       & \third{$0.8406 \pm 0.006$} \\ \midrule
SGM \cite{rogen2003automatic}                           & \multicolumn{1}{c}{-} & -                       & $0.6289$             \\ \midrule
SSEF    \cite{zotenko2006secondary}                       & \multicolumn{1}{c}{-} & -                       & $0.4920$             \\ \midrule
DeepFold   \cite{liu2018learning}                     & \multicolumn{1}{c}{-} & -                       & $0.7615$             \\ \midrule
GraSR   \cite{xia2022fast}                       & \multicolumn{1}{c}{-} & -                       & $0.8124$             \\ \bottomrule
\end{tabular}
\label{tab:tl_1}
\end{table}
Euclidean and Square geometry of the embedding space exhibited the best classification performances. Despite being trained on OOD samples, the proposed framework with Euclidean and Square geometry still managed to outperform the current state-of-the-art trained and tested on SCOPe proteins, as shown in Table~\ref{tab:tl_1}. The superior performance, despite the different training data, suggests the ability of the proposed approach to learn meaningful protein representations by 1) merging structural and sequence information into a single pipeline, and 2) capturing different and relevant properties of the geometries of the latent space into which embeddings are projected. 

\begin{figure}[h]
    \centering
    \includegraphics[width=0.85\textwidth]{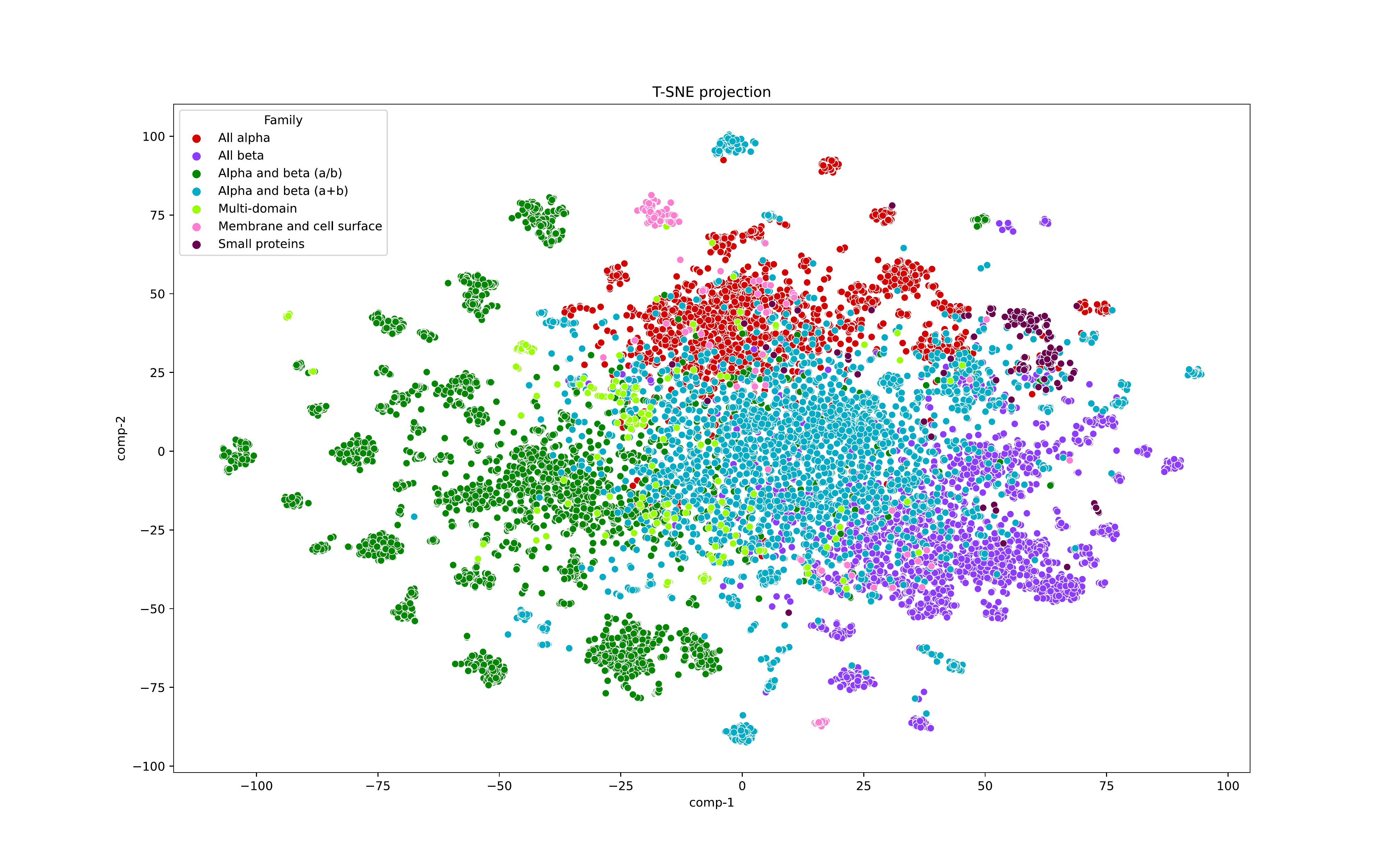}
    \caption{t-SNE visualization of the learned embeddings, coloured by protein structural family. The proposed approach generates protein embeddings which recapitulate the different families in the SCOPe dataset.
}
    \label{fig:TNSE}
\end{figure}

\paragraph{Protein structural classification} 
We constructed protein graphs with LSTM features and trained the proposed GAT architectures on the SCOPe dataset. The resulting MSE scores are reported in Appendix~\ref{MSESCOPE}. The lowest score was again reached when using Euclidean geometry for the latent space. Using this model, we projected the protein embeddings onto two dimensions using t-SNE~\cite{van2008visualizing} as shown in Figure~\ref{fig:TNSE}.
The high-level structural classes as defined in SCOPe were captured by the proposed embeddings. While not directly trained for this task, combining structural and sequence information allowed us to identify small, local clusters representing the different protein families in the SCOPe dataset. We employed supervised learning and trained a 3-layer MLP classifier to label each protein embedding in the correct family. Results of this evaluation, measured as average F1-score across 5 folds, are shown in Table~\ref{tab:tl_1} (F1-Score). When directly trained on SCOPe proteins, the proposed approach outperforms the others by a large margin for all choices of geometries (Table~\ref{tab:tl_1}).

\paragraph{Multimodal (sequence and structure) phylogenetic reconstruction} Discovering the hierarchical structure defined by evolutionary history is pivotal in numerous biological applications~\cite{de2014phylogenetic}. Inference of phylogeny is traditionally performed using sequence similarity. Here, we explore the ability of the proposed embeddings to recapitulate such hierarchy in unsupervised settings. To this end, we downloaded P53 tumour proteins from 20 different organisms from UniProt~\cite{uniprot2023uniprot} and generated protein embeddings using our GAT architecture pre-trained on SCOPe proteins. Hierarchical clustering of these embeddings, shown in Appendix \ref{phylogenetic}, demonstrates that the proposed multimodal combination results in a consistent evolutionary tree, extending phylogenetic analysis from sequence only to a combination of structure and sequence. 
%We evaluated the ability of the proposed embeddings to recapitulate such hierarchy in unsupervised settings. We downloaded P53 tumour proteins from 20 different organisms from UniProt and generate protein embeddings using our GAT architecture pre-trained on SCOPe proteins. The 

%\paragraph{Phylogenic inference from embeddings} Discovering the hierarchical structure defined by evolutionary history is pivotal in numerous biological applications. Inference of phylogeny is performed using sequence similarity. Here, we intent to explore 
%We evaluated the ability of the proposed embeddings to recapitulate such hierarchy in unsupervised settings. We downloaded P53 tumour proteins from 20 different organisms from UniProt and generate protein embeddings using our GAT architecture pre-trained on SCOPe proteins. The 

%\sbh{Table 4 should probably be split into 2 parts.}

\section{Conclusion and Future Developments}

\paragraph{Limitations} We acknowledge several limitations of our work. First, the training of the proposed models relies on the availability of structural distance scores between protein graphs. While computing these is too expensive for large-scale analysis, it is feasible to produce sufficient structural comparisons to build a training set. In addition, more complex geometries like hyperbolic~\cite{nickel2017poincare} or product spaces~\cite{gu2019learning} could be able to capture more expressive protein representations. Finally, our approach is an alignment-free method; while faster, due to the lack of structural alignment, the accuracy of alignment-free methods is generally inferior to alignment-based methods. Using deep learning architectures to extract accurate superposition information could prove beneficial for fast and more accurate protein structural comparisons. 

\paragraph{Broader Impacts} This work proposes a novel and efficient framework for comparing and analyzing protein structures. By exploiting the power of GNNs and LLMs, we generated embeddings that capture both structure and sequence information, and demonstrated that these lower-dimensional representations are useful to accurately solve downstream tasks of protein classification. A wide range of biomedical and chemical applications would benefit from the accurate representation of protein structures, such as molecular design~\cite{elton2019deep}, protein property prediction~\cite{wieder2020compact}, protein-protein and protein-ligand interaction prediction~\cite{gainza2020deciphering, yang2020graph}. Importantly, protein structures play a critical role in drug discovery, as they are often the target of drug molecules. The effectiveness of a drug and its propensity to cause unintended effects depend on how well it binds and specifically interacts with its intended target~\cite{wieder2020compact}. Measuring these factors involves expensive and time-consuming experiments. To speed up the study of drug-target interactions, learning to efficiently represent proteins is pivotal in modelling the protein targets of small molecules and their binding affinities and specificities~\cite{ma2018drug, jiang2020drug}. Several other examples of areas of application of the proposed approach are given in Appendix~\ref{BioAPP}.

\paragraph{Conclusion}
In this paper, we presented a novel framework for generating both structure- and sequence-aware protein representations. We mapped protein graphs with sequence attributes into geometric vector spaces, and showed the importance of considering different geometries of the latent space to match the underlying data distributions. We showed that the generated embeddings are successful in the task of protein structure comparison, while providing an accurate and efficient way to compute similarity scores for large-scale datasets, compared to traditional approaches (Appendix~\ref{TimeInference}). The protein graph representations generated by our approach showed state-of-the-art results for the task of protein structural classification on the SCOPe dataset. This work opens opportunities for future research, with potential for significant contributions to the fields of bioinformatics, structural protein representation and drug discovery (Appendix~\ref{BioAPP}).

\paragraph{Future Developments} While evaluated on the task of protein structure comparison, the proposed framework offers a general approach to project input graphs into geometric spaces. Assessing its capabilities when applied to small molecules (for example drugs) and omics data is left for future work. Furthermore, topological deep learning~\cite{hajijtopological} is a rapidly-emerging area of research whose success has been shown in a wide range of areas~\cite{xu2018powerful, bodnar2021weisfeiler,suk2022surfing,bodnar2022neural}. Compared to GNNs, which model the pairwise interactions on graph-structured data, several recent approaches consider more complex topological spaces such as simplicial complexes~\cite{ebli2020simplicial} or cell complexes ~\cite{hajij2020cell}. Developing new strategies for graph representation which exploit higher-order structures could prove more suitable for modelling real-world complex systems, such as protein and gene regulatory networks.

%\clearpage
\bibliographystyle{unsrtnat}
\bibliography{references}
%\clearpage
\appendix

\section{Graph Architectures}
\label{GA}

\subsection{Graph Neural Networks}

Graph Neural Networks (GNNs) are a class of neural networks that operate on data defined over graphs. Since their introduction~\citep{scarselli2008graph}, GNNs have shown outstanding results in a broad range of applications, from \emph{computational chemistry}~\citep{gilmer2017neural} to~\emph{protein folding}~\citep{jumper2021highly}. The key idea is to exploit the inductive bias induced by the topology of graph-structured data, represented by the connectivity structure, to perform graph representation learning tasks.

A graph $G=(V,E)$ is a structure that consists of a set $V$ of $n$ nodes and a set of edges $E$. In this context, each node $v \in V$ is equipped with a $d$-dimensional feature vector $\mathbf{x}_v$, and these can be grouped into a feature matrix $\mathbf{X} \in \mathbb{R}^{n \times d}$ by stacking all the $n = \vert V \vert$ feature vectors vertically. The connectivity structure of $G$ is fully captured by the adjacency matrix $\mathbf{A}$, in which the entry $i,j$ of $\mathbf{A}$ is equal to $1$ if node $i$ is connected to node $j$ and $0$ otherwise. Here, $\mathbf{A}$ is \textit{symmetric} (that is, $\mathbf{A} = \mathbf{A}^T$). In GNNs, each layer consists of a nonlinear function that maps a feature matrix into a new (hidden) feature matrix, accounting for the pairwise relationships in the underlying graph captured by its connectivity. Formally,
\begin{equation}\label{eq:gnn_general}
    \mathbf{H}^{(l)} = f(\mathbf{H}^{(l-1)}; \mathbf{A})
\end{equation}
where $\mathbf{H}^{(l)}$ is the hidden feature matrix at layer $l$ and $\mathbf{H}^{(0)} = \mathbf{X}$. Among the plethora of neural architectures that have this structure, one of the most popular is the Graph Convolutional Network~\cite{kipf2016semi}, which implements Equation \ref{eq:gnn_general} as
\begin{equation}
    \mathbf{H}^{(l)} = \sigma(\Tilde{\mathbf{D}}^{-\frac{1}{2}} \Tilde{\mathbf{A}} \Tilde{\mathbf{D}}^{-\frac{1}{2}}  \mathbf{H}^{(l-1)}\mathbf{W}^{(l)})
\end{equation}
where $\mathbf{W}^{(l)}$ is a learnable weight matrix, $\Tilde{\mathbf{A}} = \mathbf{A} + \mathbf{I}$,  $\Tilde{\mathbf{D}}$ is a diagonal matrix whose entries are $\Tilde{\mathbf{D}}_{ii} = \sum_{j}{\Tilde{\mathbf{A}}_{ij}}$ and $\sigma$ is a pointwise nonlinear activation function (for example, Sigmoid, Tanh, ReLU).

\subsection{Graph Attention Network}

The Graph Attention Network (GAT)~\cite{velivckovic2017graph} is a type of GNN that uses attention mechanisms to capture dependencies between nodes in a graph. The key idea behind GATs is to learn a different weight for each neighboring node in the graph using a shared attention mechanism. This allows a GAT to attend to different parts of the graph when computing the representation of each node. The GAT layer can be mathematically expressed as
\begin{equation}
h_i^{(l+1)} = \sigma\left(\sum_{j\in\mathcal{N}_i}\alpha_{ij}^{(l)}\mathbf{W}^{(l)}\mathbf{h}_j^{(l)}\right)
\end{equation}
where $\mathbf{h}_i^{(l)}$ denotes the representation of node $i$ at layer $l$, $\mathcal{N}_i$ represents the set of neighbouring nodes of $i$, $\alpha_{ij}^{(l)}$ represents the attention weight between nodes $i$ and $j$ at layer $l$, $\mathbf{W}^{(l)}$ is the weight matrix at layer $l$, and $\sigma$ is the activation function. The coefficients computed by the attention mechanism can be expressed as:
\begin{equation}
\alpha_{ij} = \frac{\exp\left(\mathrm{LeakyReLU}\left(\mathbf{a}^{\top}[\mathbf{W}^{(l)}\mathbf{h}_i^{(l)} || \mathbf{W}^{(l)}\mathbf{h}_j^{(l)}]\right)\right)}{\sum_{k\in\mathcal{N}_i}\exp\left(\mathrm{LeakyReLU}\left(\mathbf{a}^{\top}[\mathbf{W}^{(l)}\mathbf{h}_i^{(l)} || \mathbf{W}^{(l)}\mathbf{h}_k^{(l)}]\right)\right)}
\end{equation}
where $[\cdot || \cdot]$ denotes concatenation, $\mathbf{a}^{\top}$ is a trainable weight vector, and $\mathrm{LeakyReLU}$ is the Leaky Rectified Linear Unit activation function. 

%\sbh{You need to use a uniform notation. With vectors and matrices, you have a mixture of italic, bold and over-arrow. Don't mix them like this --- decide on one and use it everywhere.}

\subsection{GraphSAGE}

GraphSAGE~\cite{hamilton2017inductive} is a type of GNN that learns node representations by aggregating information from the local neighborhood of each node. GraphSAGE learns a set of functions to aggregate the representations of a node's neighbors, and then combine them with the node's own representation to compute its updated representation. The GraphSAGE layer can be mathematically expressed as
\begin{equation}
h_i^{(l+1)} = \sigma\left(\mathbf{W}^{(l)}\cdot\mathrm{CONCAT}\left(\mathrm{AGGREGATE}\left({\mathbf{h}_j^{(l)}: j\in\mathcal{N}_i}\right),\mathbf{h}_i^{(l)}\right)\right)
\end{equation}
where $\mathbf{h}_i^{(l)}$ denotes the representation of node $i$ at layer $l$, $\mathcal{N}_i$ represents the set of neighbouring nodes of $i$, $\mathrm{AGGREGATE}$ is a learnable aggregation function that combines the representations of a node's neighbors, $\mathrm{CONCAT}$ is the concatenation operation, $\mathbf{W}^{(l)}$ is the weight matrix at layer $l$, and $\sigma$ is the activation function. 

%\sbh{Now you also have two different notations for concatenation.}

\section{Distance Functions}
\label{DF}

The proposed approach is to map graphs into a continuous space so that the distance between embedded points is correlated to the distance between the original graphs measured by the TM-score. We explored different distance functions in the embedding space, and we give here their definitions. Given a pair of vectors $\textbf{p}$ and $\textbf{q}$ of dimension $k$, the definitions of the Manhattan, Euclidean, Square and Cosine distances are as follows:
\begin{equation}
    \text{Manhattan: } d( \textbf{p},  \textbf{q}) = \|\textbf{p}-\textbf{q}\|_1 = \sum_{i=0}^k|p_i - q_i|
\end{equation}
\begin{equation}
    \text{Euclidean: } d( \textbf{p},  \textbf{q}) = \|\textbf{p}-\textbf{q}\|_2 = \sqrt{\sum_{i=0}^k(p_i - q_i)^2}
\end{equation}
\begin{equation}
    \text{Square: } d( \textbf{p},  \textbf{q}) = \|\textbf{p}-\textbf{q}\|_2^2 = \sum_{i=0}^k(p_i - q_i)^2
\end{equation}
\begin{equation}
    \text{Cosine: } d( \textbf{p},  \textbf{q}) = 1 - \frac{ \textbf{p}  \cdot \textbf{q}}{\|\textbf{p}\|\|\textbf{q}\|} = 1 - \frac{\sum^k_{i=0}p_iq_i}{\sqrt{\sum^k_{i=0}p_i^2}\sqrt{\sum^k_{i=0}q_i^2}} .
\end{equation}

\section{Datasets}

\paragraph{Kinase proteins}
We downloaded the human proteome from UniProt\footnote{\url{https://www.uniprot.org}} and sub-selected 512 protein kinases. We also used UniProt to download the PDB files for the kinases. 

\paragraph{SCOPe v2.07}
The 40\% identity filtered subset of SCOPe v2.07\footnote{\url{https://scop.berkeley.edu/help/ver=2.07}} is used to train and validate our approach. Out of the total of 14,323 domains, 1,058 domains were removed during the data collection process. The remaining 13,265 domains were used for training and testing. 
For both datasets, we computed ground truth TM-scores by performing all-against-all comparisons using TM-align~\cite{zhang2005tm}. We used 80\% of the comparisons for training and 20\% for testing. We repeated all the experiments with 3 different seeds. 

\section{Additional experiments and details}
\label{AED}

\subsection{TM-scores predictions}
\label{TMAPP}

We employed the trained GAT architectures from Table~\ref{tab:res1_table} to predict the TM-scores for the kinase pairs in the test set. Results of this evaluation, measured by Pearson correlation between model predictions and true TM-scores, are shown in Table~\ref{tab:corr_coeff}. 
\begin{table}[h!]
\centering
\caption{Pearson correlation coefficients between predicted and actual TM-scores for the GAT model for different choices of node features and distance functions. We use gold \tikzcircle[gold,fill=gold]{2pt}, silver \tikzcircle[silver,fill=silver]{2pt}, and bronze \tikzcircle[bronze,fill=bronze]{2pt} colors to indicate the first, second and third best performances, respectively. The highest score is reached with LSTM-extracted features and Euclidean geometry. }
\vspace{3px}
\begin{tabular}{@{}lcccc@{}}
\toprule
\multicolumn{1}{l}{\textbf{Feature}} & \multicolumn{4}{c}{\textbf{Distance}}            \\
\multicolumn{1}{l}{} & \texttt{Cosine} & \texttt{Euclidean} & \texttt{Manhattan} & \texttt{Square} \\ \midrule
One-Hot              & $0.661$  & $0.384$     & $0.637$     & $0.226$  \\
Physicochemical      & $0.463$  & $0.358$     & $0.534$     & $0.166$  \\
BLOSUM               & $0.484$  & $0.658$     & $0.761$     & $0.468$  \\
BERT features        & $0.849$  & \second{$0.870$}      & $0.837$     & $0.785$  \\
LSTM features        & \third{$0.861$}  & \first{$0.879$}     & $0.858$     & $0.839$  \\ \bottomrule
\end{tabular}

\label{tab:corr_coeff}
\end{table}
Using features learned by LLMs exhibits superior performance compared to other feature extraction methods. The highest score is reached with LSTM-extracted features and Euclidean geometry of the embedding space.   

\subsection{TM-scores inference times}
\label{TimeInference}

Table~\ref{tab:inf_time} 
\begin{table}[]
\centering
\caption{Wall-clock estimates for the GNN models and TM-align on different percentages of the test set. Among the GNNs, GAT is the slowest at computing TM-scores, followed by GraphSAGE and GCN, both on GPU and CPU. However, TM-score computation with any of the GNN architectures is significantly faster than TM-align, even on CPU.}
\vspace{3px}
\label{tab:inf_time}
\begin{tabular}{@{}clcc@{}}
\toprule
\textbf{Test Size (\%)}       & \textbf{Model} & \textbf{GPU Inference (s)} & \textbf{CPU Inference (s)} \\ \midrule
\multirow{4}{*}{26164 (20\%)} & GCN            & $88.3 \pm 2.04$            & $474.78 \pm 1.98$          \\
                              & GAT            & $125.8 \pm 2.26$           & $1570.1 \pm 3.21$          \\
                              & GraphSAGE      & $ 98.2 \pm 3.46$           & $618.2 \pm 2.56$           \\
                              & TM-align       & -                          &        $57659.3 \approx 16 \text{ hr}$                   \\ \midrule
\multirow{4}{*}{13082 (10\%)} & GCN            & $49.2 \pm 0.53$            & $231.1 \pm 3.01$           \\
                              & GAT            & $59.3 \pm 2.34$            & $773.6 \pm 3.14$           \\
                              & GraphSAGE      & $49.3 \pm 0.04$            & $313.9 \pm 2.23$           \\
                              & TM-align       & -                          &       $29156.2 \approx 8 \text{ hr}$                  \\ \midrule
\multirow{4}{*}{6541 (5\%)}   & GCN            & $23.2 \pm 0.18$            & $119.6 \pm 1.76$           \\
                              & GAT            & $30.1 \pm 0.70$            & $3882 \pm 3.26$            \\
                              & GraphSAGE      & $25.6 \pm 1.42$            & $153.1 \pm 3.01$           \\
                              & TM-align       & -                          &       $15019.9 \approx 4 \text{ hr}$                    \\ \bottomrule
\end{tabular}
\end{table}
provides inference times for the different graph models (GCN, GAT and GraphSAGE) and TM-align~\cite{zhang2005tm}. We show the inference times on GPU and CPU for the graph models, and CPU time for TM-align. Time estimates for different percentages of the test set (20\%, 10\%, 5\%) are reported. For the graph models, we also report standard deviations by estimating the times over 5 different runs. The GNN architectures are significantly faster than TM-align, even on CPU. Our approach represents a fast (Table~\ref{tab:inf_time}) and accurate (Table~\ref{tab:corr_coeff}) way to compute protein structural similarities even on large-scale datasets.  

\subsection{MSE results on SCOPe proteins}
\label{MSESCOPE}

Table~\ref{tab:scope_MSE} 
\begin{table}[h]
\centering

\caption{MSE scores for different distance functions and LSTM features on the SCOPe dataset. We use gold \tikzcircle[gold,fill=gold]{2pt}, silver \tikzcircle[silver,fill=silver]{2pt}, and bronze \tikzcircle[bronze,fill=bronze]{2pt} colors to indicate the first, second and third best performances, respectively.}
\begin{tabular}{@{}lll@{}}
\toprule
\textbf{Model}&\textbf{Distance} & \textbf{MSE}      \\ \midrule
\multirow{4}{*}{ GAT} & \texttt{Cosine}            & \second{$0.008048$} \\
&\texttt{Euclidean}         & \first{$0.006294$} \\
&\texttt{Manhattan}        & $0.010655$\\
&\texttt{Square}            & \third{$0.008793$} \\ \bottomrule
\end{tabular}
\label{tab:scope_MSE}
\end{table} 
reports the MSE scores for different distance functions and LSTM features on the SCOPe dataset. The best MSE is again reached with LSTM-extracted features and Euclidean geometry of the embedding space.

\subsection{Computational Resources and Code Assets} 

In all experiments we used NVIDIA\textsuperscript{\textregistered} Tesla V100 GPUs with 5,120 CUDA cores and 32GB GPU memory on a personal computing platform with an Intel\textsuperscript{\textregistered} Xeon\textsuperscript{\textregistered} Gold 5218 CPU @ 2.30GHz CPU using Ubuntu 18.04.6 LTS. The models have been implemented in PyTorch~\citep{NEURIPS2019_9015} and PyTorch Geometric library\footnote{\url{https://github.com/pyg-team/pytorch_geometric/}}~\cite{pyg2019}. PyTorch, NumPy, SciPy are made available under the BSD license, Matplotlib under the PSF license, and PyTorch Geometric is made available under the MIT license.

\section{Phylogenetic reconstruction}
\label{phylogenetic}

\begin{figure}[]
    \centering
    \includegraphics[width=0.75\textwidth]{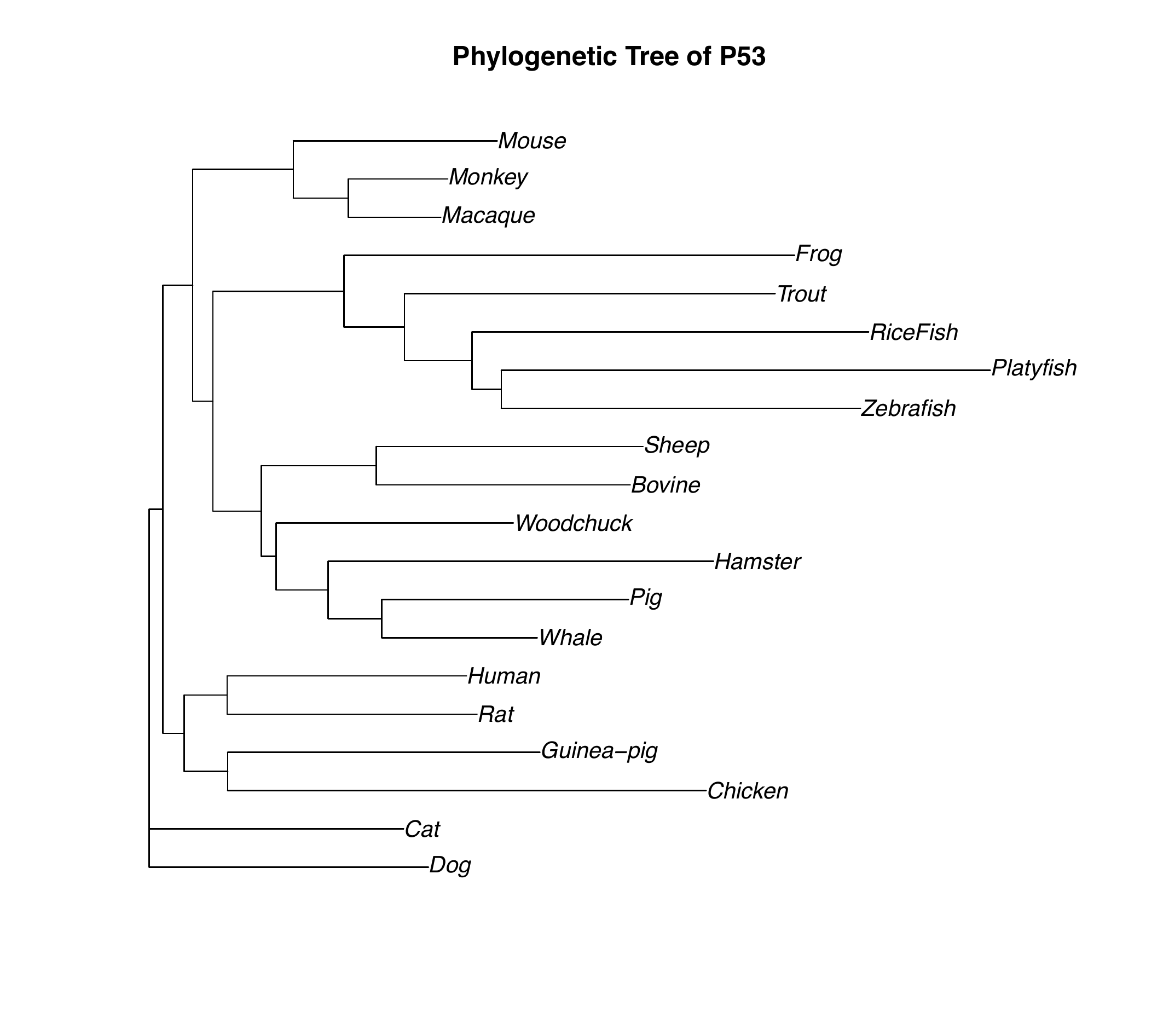}
    \caption{Phylogenetic tree for P53 tumor protein across 20 species. }
    \label{fig:phylo}
\end{figure}

Phylogenetic reconstruction refers to the task of inferring evolutionary relationships based on morphological or physiological characters. Typical phylogenetic reconstruction makes use of sequence similarity only. Here, we explore the ability of the proposed embeddings to recapitulate phylogenies in unsupervised settings. To this end, we downloaded P53 tumour proteins from 20 different organisms from UniProt~\cite{uniprot2023uniprot} and generated protein embeddings using our GAT architecture pre-trained on SCOPe proteins. A common method for Tree Estimation, Neighbor-Joining, was used to build the phylogenetic tree, shown in Figure \ref{fig:phylo}.
The proposed embeddings managed to accurately recapitulate the evolutionary relationships among the 20 species (Figure \ref{fig:phylo}). Our embeddings (structure- and sequence-aware) offer an extension to traditional methods for phylogenetic reconstruction from sequence only to a combination of structure and sequence.

\section{Bioinformatics applications}
\label{BioAPP}

There are several areas of bioinformatics research where structural representation of proteins finds useful applications. We now give a few examples. 

\paragraph{Protein-protein interaction} Proteins rarely carry out their tasks in isolation, but interact with other proteins present in their surroundings to complete biological activities. Knowledge of protein–protein interactions (PPIs) helps unravel cellular behaviour and functionality. Generating meaningful representations of proteins based on chemical and structural information to predict protein-pocket ligand interactions and protein-protein interactions is an essential bioinformatics task~\cite{yang2020graph}.

\paragraph{Protein function} The structural features of a protein determine a wide range of functions: from binding specificity and conferring of mechanical stability, to catalysis of biochemical reactions, transport, and signal transduction. While the experimental characterization of a protein's functionality is a challenging and intense task~\cite{moreau2012computational}, exploiting graph representation learning ability to incorporate structural information facilitates the prediction of protein function~\cite{zhou2021functions, gligorijevic2021structure,zhao2022panda2}.

\paragraph{Small molecules} The design of a new drug requires experimentalists to identify the chemical structure of the candidate drug, its target, its efficacy and toxicity and its potential side effects~\cite{hu2016network, rai2018network,barabasi2011network}. Because such processes are costly and time consuming, drug-discovery pipelines employ in silico approaches. Effective representations of protein targets of small molecules (drugs) has the potential to dramatically speed up the field of drug discovery. 

\end{document}